\documentclass[prl,twocolumn,showpacs,preprintnumbers,amsmath,amssymb]{revtex4}

\usepackage{graphicx}
\usepackage{dcolumn}
\usepackage{bm}

\begin{document}

\title{Glassy dynamics, spinodal fluctuations, and the kinetic limit of hard-rod nucleation}

\author{Ran Ni$^1$, Simone Belli$^2$, Ren{\'e} van Roij$^2$, and Marjolein Dijkstra$^1$}
\affiliation{$^1$Soft Condensed Matter,  Utrecht University, Princetonplein 5, 3584 CC Utrecht, The Netherlands}
\affiliation{$^2$Institute for Theoretical Physics, Utrecht University, Leuvenln 4, 3504 CE Utrecht, The
Netherlands}

\date{\today}

\begin{abstract}
Using  simulations we identify three dynamic regimes in supersaturated isotropic fluid states of short hard rods:
(i) for moderate supersaturations we observe nucleation of multi-layered crystalline clusters; (ii) at higher
supersaturation, we find nucleation of small crystallites which arrange into long-lived locally favored structures
that  get kinetically arrested, while (iii) at even higher supersaturation the dynamic arrest is due to the
conventional cage-trapping glass transition. For longer rods we find that the formation of the (stable) smectic
phase out of a supersaturated isotropic state is strongly suppressed by an isotropic-nematic spinodal instability
that causes huge spinodal-like orientation fluctuations with nematic clusters diverging in size. Our results show
that glassy dynamics and spinodal instabilities set kinetic limits to nucleation in a highly supersaturated hard-rod fluids.
\end{abstract}

\maketitle
Nucleation is the process whereby a thermodynamically metastable state evolves into a stable one, via the spontaneous formation of a
droplet of the stable phase. According to classical nucleation theory (CNT), the Gibbs free energy associated with the formation of a spherical
cluster of the stable phase with radius $R$ in the metastable phase is given by a volume term, which represents  the driving force to form the new
phase, and a surface free energy cost to create an interface, i.e., $\Delta G = -4 \pi R^3 \rho |\Delta\mu|/3 + 4 \pi R^2 \gamma$ with $\gamma$
the surface tension between the coexisting phases, $\rho$ the  density of the cluster, and $|\Delta\mu|>0$ the chemical potential difference
between the metastable and stable phase. For a given $|\Delta \mu|$ and $\rho$, CNT predicts a nucleation barrier $\Delta G_{crit} = (16 \pi /3)
\gamma^3/(\rho|\Delta \mu|)^2$ and a critical nucleus radius $R_{crit}=2 \gamma/\rho|\Delta \mu|$. CNT predicts an infinite barrier at bulk
coexistence ($\Delta\mu=0$), which decreases with increasing supersaturation. However, CNT incorrectly predicts a {\em finite} barrier at the
spinodal, whereas a non-classical approach yields a vanishing barrier at the spinodal, with a diffuse critical nucleus that becomes of infinite
size \cite{cahn}. Both approaches explain why liquids must be supercooled substantially before nucleation occurs, and one might expect that
nucleation should always occur for sufficiently high supersaturation.
For deep quenches of soft spheres close to a spinodal, but not beyond it, simulation studies show either nucleating anisotropic and diffuse clusters
\cite{klein}, or precritical clusters that grow further \cite{parrinello} or that coalesce in ramified structures  \cite{bagchi}. These results
contrast the mean-field predictions that the critical size should diverge at the spinodal \cite{cahn}. On the other hand, Wedekind {\em et al.}
showed that a Lennard-Jones system can become unstable by a so-called kinetic spinodal, where the largest cluster in the system has a vanishing barrier, i.e.
$\Delta G_{crit}^{large}=0$, implying the immediate formation of a critical cluster in the system  \cite{wedekind}. Beyond this kinetic limit,
which is system-size dependent as $\Delta G_{crit}^{large}=\Delta G_{crit}-k_BT\ln N$, the system is kinetically unstable, and the phase
transformation proceeds immediately via growth of the largest cluster. Here $N$ is the number of particles, $k_B$ the Boltzmann constant, and $T$
the temperature. This scenario also explains why it is hard to reach the thermodynamic spinodal and why a divergence of the critical cluster size
is never observed in simulations, as the system already becomes kinetically unstable at much lower supersaturations. Interestingly, recent
simulations of silica also showed a kinetic limit of the homogeneous nucleation regime that is strongly influenced by glassy dynamics, without any
spinodal effects \cite{poole}. Clearly,  the nucleation kinetics at high supersaturation is still poorly understood.

\begin{figure*}
\includegraphics[width=0.9\textwidth]{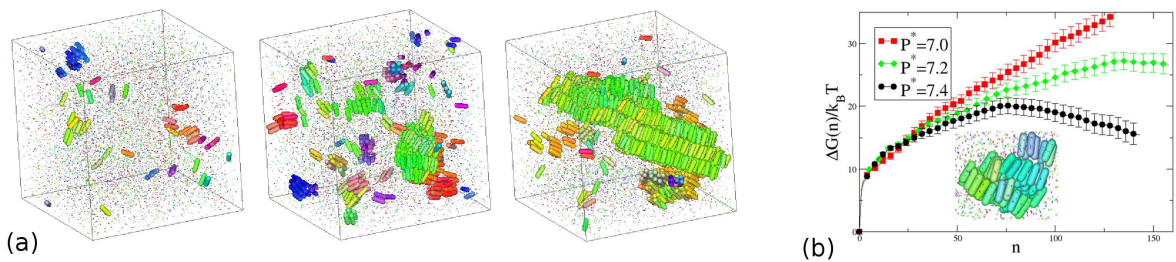}
\caption{\label{fig:snapshot76}(color online) (a) Configurations for spontaneous
crystal nucleation from a typical molecular dynamics trajectory at
$P^*=7.6$ and $t/\tau=0,1000$ and $3000$ (from left to
right) with $\tau=\sigma\sqrt{m/k_BT}$ and $m$ the mass of the
particle. Isotropic-like particles are drawn $10$ times smaller
than their actual size. A movie can be found in \cite{supplementary}. (b) Gibbs free energy $\Delta G(n)$ as a
function of the number of rods $n$ in the crystalline cluster at
pressure $P^*= 7.0, 7.2,$ and 7.4. Inset: A typical configuration of a
critical cluster
($n=81$) at $P^*=7.4$.}
\end{figure*}

In this Letter, we investigate not only the nucleation pathways of the isotropic-crystal (IX) transition of
rod-like particles as a function of supersaturation, but also those of the isotropic-smectic (ISm) transition. The
nucleation pathways  of structures with both orientational and positional order are still unknown, as nucleating
smectic or crystalline clusters have never been observed in experiments or simulations \cite{schilling2004,patti}.
We show for the first time that crystal nucleation proceeds via nucleation of multi-layer crystalline clusters,
while previous studies found that nucleation is hampered by self-poisoning \cite{schilling2004}. Additionally, we
identify two mechanisms of dynamic arrest that sets a kinetic limit to the crystal nucleation regime, one based
on dynamic arrest of small crystalline nuclei that form locally favored structures, and one based on a
conventional cage-trapping glass transition. Moreover, for longer rods we show that the 
isotropic-nematic (IN) spinodal associated with a metastable IN transition severely hinders and even prevents ISm nucleation.

We consider a suspension of $N$ hard spherocylinders  with a diameter $\sigma$ and a cylindrical segment of length $L=2\sigma$ in a volume $V$ or
at pressure $P$. The  bulk phase diagram of these rods with a length-to-diameter ratio $L^*=L/\sigma = 2$ is well known
\cite{peter1997}; it features an IX  transition at pressure $P^*=\beta P\sigma^3=5.64$  with $\beta=1/k_BT$.

We first use {\it NPT}-Monte Carlo (MC) simulations to compress an isotropic fluid of $10,000$ rods at the moderate pressure $P^*=7.6$
corresponding to a chemical potential difference $\beta |\Delta \mu| = 1.11$ between the (metastable) fluid and the crystal phase. We then take
random MC configurations as initial configurations for molecular dynamics (MD) simulations in the  {\it NVT} ensemble to study spontaneous crystal
nucleation, employing the cluster criterion as described in Ref.~\cite{cuetos2007,supplementary}. We find spontaneous nucleation of a
multi-layered crystalline cluster in the isotropic fluid. Fig.~\ref{fig:snapshot76}a shows the time evolution from a typical MD trajectory. In the
initial stage of the MD simulation the system remains in the metastable isotropic fluid for a long time, with small multi-layered crystalline
clusters appearing and disappearing along the simulation. After time $t=1000\tau$, with time unit $\tau=\sigma\sqrt{m/k_BT}$ and $m$ the mass of
the particle, a nucleus consisting of multiple crystalline layers  starts to grow gradually until the whole system has been transformed into the
bulk crystal phase.  We note that the cluster prefers to grow laterally as was also found for attractive rods \cite{patti}. We observed similar
spontaneous nucleation at $P^*=7.4$. The long waiting time $t_w$ before a postcritical cluster starts to appear by a spontaneous fluctuation is
typical for nucleation and growth.  We calculate the nucleation rate $R=1/\langle t_w \rangle V$, and find from our MD simulations that $R=5\times
10^{-9\pm2}\tau^{-1}\sigma^{-3}$ and $1.7\times 10^{-8\pm1}\tau^{-1}\sigma^{-3}$, for $P^*=7.4$ and 7.6, respectively.

As our MD simulations provide evidence that the IX transformation can occur via nucleation of multilayer crystalline clusters, we determine the
nucleation barrier using umbrella sampling (US) in MC simulations.  We bias the system to configurations with a certain cluster size and we sample
the equilibrium probability $P(n)$ to find a cluster of $n$ rods. The Gibbs free energy of a cluster of size $n$ is then given by $\beta \Delta
G(n)=-\mathrm{ln}P(n)$. We perform MC simulations of  $2000$ particles at $P^* = 7.0, 7.2$, and $7.4$ corresponding to $\beta |\Delta \mu| = 0.78,
0.89$ and $1.0$, respectively. Fig.~\ref{fig:snapshot76}b shows $\Delta G(n)$, which for $P^*=7.2$ and $7.4$ display a maximum of $\beta \Delta
G_{crit} \approx 27\pm1.5$ and $20\pm1.5$ at critical cluster sizes $n_{crit}\approx 140$ and $80$, respectively. A typical configuration of the
critical cluster, consisting of three crystalline layers at $P^*=7.4$, is shown in the inset of Fig.~\ref{fig:snapshot76}b; its structure agrees
with those observed in our MD simulations of spontaneous nucleation of multilayer crystallites. For $P^*=7.0$  the free-energy barrier is too high
to be calculated in our simulations as the cluster starts to percolate the simulation box before the top is reached. For even lower pressures,
i.e. $P^*=6.0$ (not shown), this problem is even more severe. For clusters up to $n\simeq 100$, however, the barrier can be calculated with the US
scheme, revealing multilayered structures very similar to the one shown for $P^*=7.4$.
Our MC simulation results for $P^*=7.2$ and $7.4$ can also be used to calculate the nucleation rate from $R=\kappa \exp\left(-\beta\Delta
G_{crit}\right)$ with kinetic prefactor $\kappa= \left|\beta\Delta G''_{crit}/(2 \pi)\right|^{1/2} \rho_I f_{n_{crit}}$, where   $\rho_I$ is the
number density of the isotropic fluid and $f_{n_{crit}}$ is the attachment rate of particles to the critical cluster (which we compute using MD
simulations starting with independent configurations at the top of the nucleation barrier ~\cite{auer2001}). For $P^*=7.2$ and $7.4$ we find
$R=1\times 10^{-13\pm 1}$ and $2\times 10^{-10\pm 1}$ $\tau^{-1}\sigma^{-3}$, respectively,  in agreement within errorbars with the MD
simulations.

Our observation of spontaneous nucleation of bulk crystals of short rods is in marked contrast with an earlier study, which showed that the free
energy never crosses a nucleation barrier~\cite{schilling2004}. These simulations showed the formation of a single crystalline layer, while
subsequent crystal growth is hampered. The authors attributed the stunted growth of this monolayer to self-poisoning by rods that lie flat on the
cluster surface. If we use the same cluster criterion as in Ref.~\cite{schilling2004} for the biasing potential, we indeed also find crystalline
monolayers at $P^*=7.4$, which cannot grow further as $\Delta G(n)$ increases monotonically with $n$. These results for the nucleation barrier
agree with theoretical predictions that for sufficiently low supersaturations $\Delta G(n)$ for a single layer is always positive, while
multilayer crystalline clusters can grow spontaneously when the nucleus exceeds the critical size \cite{frenkel2002}. However, our detailed check
~\cite{supplementary} of the order parameter in Ref.~\cite{schilling2004} actually reveals a strong (unwanted) bias to form single-layered clusters
in US simulations.

We also study the IX transformation at higher supersaturation. To this end, we compress $1000$ rods ($L^*=2$) in
{\it NPT}-MC simulations at $P^*=8$ ($\beta |\Delta \mu|=1.33$). Using $\beta \gamma \sigma^2 \simeq 0.44$, which
follows from fitting the two barriers of Fig. \ref{fig:snapshot76}b to CNT, we estimate barriers as low as
$\beta\Delta G_{crit}\sim 12$ and $\beta\Delta G_{crit}^{large}\sim 5$ for $P^*=8$. Indeed, many small
crystallites nucleate immediately after the compression quench, indicative of the proximity of a kinetic spinodal.
These crystallites are oriented in different directions, and have a large tendency to orient perpendicular to each other.
The subsequent equilibration is extremely slow, since the growth of a single crystal evolves via collective
re-arrangements of smaller clusters that subsequently coalesce. In fact, after $3\times10^{7}$ MC cycles, our
system is dynamically arrested. Interestingly, Frank proposed more than 50 years ago that dynamic arrest may be
attributed to the formation of {\em locally favored structures} in which the system gets kinetically trapped in
local potential-energy minima \cite{frank}, while direct observation of such a mechanism for dynamic arrest was
only recently reported in the gel phase of a colloid-polymer mixture \cite{royall}. In our simulations, we
clearly observe the formation of long-lived locally favored structures consisting of perpendicularly oriented
crystallites. Only via cooperative rearrangements (rotation of the whole cluster) the system can escape from the
kinetic traps, but these events are rare in MC simulations. So despite the large supersaturation and the low barrier as predicted by
CNT,  the actual formation of a single crystal is impeded dramatically by slow dynamics. In fact, our observations
agree with experiments on soft-repulsive selenium rods, where transient structures of 5-10 aligned particles tend
to form locally favored structures with perpendicularly oriented clusters, which gradually merge into larger
clusters~\cite{maeda2003}. Only attractive $\beta$-FeOOH rods form crystalline monolayers in agreement with
\cite{patti}.

\begin{figure}
\includegraphics[width=0.5\textwidth]{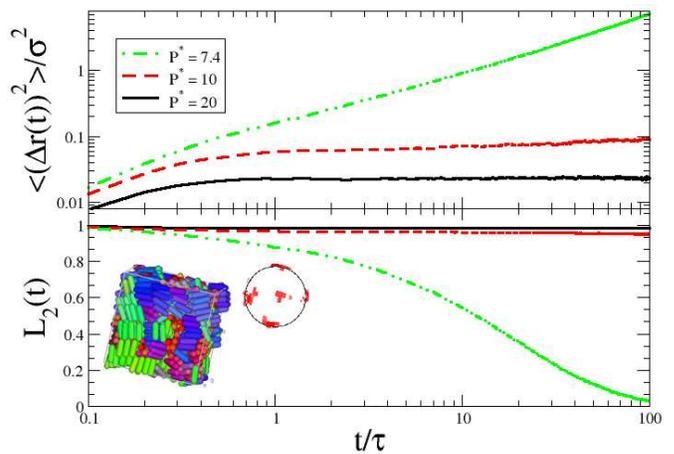}
\caption{\label{fig:spon8}(color online)
Mean square displacement $\langle (\Delta {\bf r}(t))^2 \rangle$ and second-order orientational correlator $L_2(t)$ 
for hard rods with $L^*=2$ and pressures as labeled. The inset shows a typical configuration of a glassy state with cubatic order at $P^*=10$. }
\end{figure}

In order to investigate whether the system can be quenched {\em beyond} a thermodynamic spinodal (such that the
transformation should proceed via spinodal decomposition), we also perform simulations at $P^*=10$. We find again
the immediate nucleation of many small crystallites, as expected beyond the kinetic spinodal. As the phase
transformation sets in right away, we cannot determine whether the nucleation barrier is finite or zero; it is
therefore unclear whether or not we have crossed a thermodynamic spinodal (if there is one for freezing). We note,
however, that we did not find any characteristics of early-stage spinodal decomposition. The small
crystallites tend to orient perpendicularly, and in fact the system displays clear orientational ordering along three
perpendicular directions (cubatic order), as shown by the orientation distribution on the surface of a unit sphere
in the inset of Fig.~\ref{fig:spon8}. To check for finite size effects, we studied a system of $N=4000$ rods, which again shows system-spanning  cubatic order. Whether or not  the cubatic order is long-ranged for even larger systems remains unsettled. The mean-square displacement $\langle (\Delta {\bf r}(t))^2 \rangle$ and the
second-order orientational correlator $L_2(t) = \langle (3 \cos^2 \theta(t) -1)/2 \rangle$ are also displayed in
Fig.~\ref{fig:spon8}, which show  the characteristic plateau of  structural arrest. For comparison, we also
present data for $P^*=7.4$, which show relatively fast relaxation of the translational and orientational degrees
of freedom. At an even larger supersaturation, $P^*=20$, we find that  the system is kinetically arrested
immediately after the quench. We find hardly any crystalline order, while the orientation distribution remains
isotropic (not shown). Clearly, the system crossed the conventional cage trapping glass transition \cite{letz}
that prevents the formation of any ordering. The dynamic arrest can be appreciated by the plateau in $\langle
(\Delta {\bf r}(t))^2 \rangle$ and $L_2(t)$ in Fig. \ref{fig:spon8}. Our results thus show that nucleation at high
supersaturation is strongly affected by vitrification, either due to locally favored structures or by the
conventional glass transition, yielding glasses with and without small crystallites, respectively.

\begin{figure}
\includegraphics[width=0.5\textwidth]{Fig3a.eps}
\includegraphics[width=0.5\textwidth,angle=0]{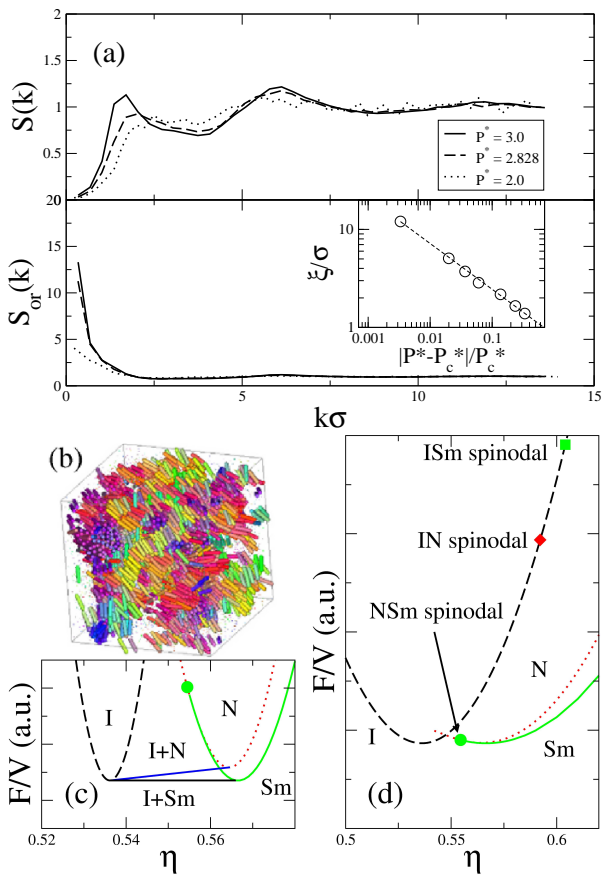}
\caption{\label{fig:msdL3p4} (color online) (a) Positional (top) and orientational (bottom) structure factor of  hard rods  with
$L/\sigma=3.4$ at varying $P^*$.  The inset shows the pressure-dependence of the orientation correlation
length $\xi$. The dashed line is the power-law fitting  $\xi \sim |P-P_c|^{-\nu}$ with $\nu=0.47$ and
$P^*_c=3.01$. (b) Typical configuration  at $P^*=3$. A movie is shown in Ref.
\cite{supplementary}. Isotropic-like particles are drawn 10 times smaller than their actual size. (c,d) 
Conveniently shifted and scaled Helmholtz free energy density $F/V$ of Zwanzig rods (packing
fraction $\eta$, aspect ratio $H/D=4.3$) in the I (dashed), N (dotted), and Sm (green) phase. 
ISm and metastable IN coexistence are indicated by the solid lines (black and blue, respectively), 
and the IN  and ISm spinodal  instabilities are denoted by the symbols (red diamond and green square, respectively) on the
supersaturated isotropic free-energy branch.
}
\end{figure}

We also study longer hard rods with $L^*=3.4$, which show  ISm coexistence at $P^*=2.828$. A previous MC simulation study \cite{cuetos2010} indeed
showed the formation of the smectic phase out of the highly supersaturated I phase at $P^*=3.1$ via spinodal decomposition. However, nucleation
and growth of the smectic phase out of weakly supersaturated I phases at $P^*=2.85-3.0$ was {\em not} observed \cite{cuetos2010}. As strong
pre-smectic ordering and huge nematic-like clusters were observed in the I phase, the hampered nucleation was attributed to slow
dynamics. Here we reinvestigate the regime $P^*=2.828-3.0$ at much longer time scales by MD simulations. We confirm the earlier findings as
regards the structure, but did not find any evidence for structural arrest in  $\langle (\Delta {\bf r}(t))^2 \rangle$ and $L_2(t)$ (not shown).
Instead we find huge and strongly fluctuating nematic-like clusters  \cite{supplementary}. The nematic character of the clusters is evident from
the structure factor $S(k)$ and orientational structure factor $S_{or}(k)$, shown in Fig.~\ref{fig:msdL3p4}, revealing a small-$k$ divergence for
$S_{or}(k)$ but not for $S(k)$ \cite{letz}. The correlation length $\xi$ of the orientational fluctuations obtained from  fitting the orientational correlation function $g_{or}(r) \sim \exp(-r/\xi)/r $ is shown in the inset to satisfy a
power law $\xi\sim |P-P_c|^{-\nu}$ with $P^*_c=3.01$ the alleged IN spinodal pressure and ${\nu}=0.47$, which is close to the expected mean-field
exponent $\nu=1/2$ of the IN-spinodal \cite{gramsbergen}. Apparently, the ISm nucleation is prevented by an intervening
 IN spinodal. Our observation that the metastable isotropic fluid is more susceptible to nematic than to smectic
fluctuations is corroborated by second-virial calculations of the Zwanzig model of block-like $H\times D\times D$ rods with three orthogonal
orientations \cite{supplementary}. The dimensionless Helmholtz free-energy density $F/V$ of the I, N, and Sm phase for $H/D=4.3$, shown in
Fig.\ref{fig:msdL3p4}, reveal equilibrium ISm coexistence, and a metastable N branch. Moreover,  the IN spinodal on the metastable isotropic branch occurs
at a lower packing fraction $\eta$ than the ISm spinodal. In other words, the isotropic fluid is predicted to exhibit spinodal nematic
fluctuations upon increasing the supersaturation, consistent with the diverging $\xi$ as observed in our simulations. One might have expected that
the  presence of these nematic clusters facilitate the formation of the smectic phase. However, although we do find some
layering of the rods, the density within these nematic clusters is too low and the orientational fluctuations change too rapidly to form the
smectic layers.

In conclusion, our results show that nucleation of short hard rods from a supersaturated isotropic fluid phase to crystal and smectic phases 
is much more rare than perhaps naively anticipated. Only for very short rods and moderate supersaturations, we
find, for the first time, nucleation of multi-layered crystals; at higher supersaturations we identified two mechanism for dynamic arrest. The
first one occurs close to the kinetic spinodal, where (locally favored) crystalline clusters  appear immediately after the quench,
followed by  slow dynamics due to geometric constraints of these tightly packed clusters. The second type of dynamic arrest occurs at
very high supersaturation and is due to the conventional cage-trapping glass transition.  In the supersaturated isotropic state of slightly 
longer rods ($L^*=3.4$), the nucleation of the (equilibrium)
smectic phase is found to be hampered by huge nematic fluctuations due to the existence of an  IN spinodal instability. In fact, we
showed for the first time that for quenches close to a spinodal the clusters diverge in size.

Our findings are of fundamental and practical interest. They provide strong evidence for a local structural
mechanism for dynamic arrest in a system with orientational and positional degrees of freedom. They also explain
why the self-organization of ordered assemblies of nanorods is difficult and why most of the nanorod self-assembly
techniques require additional alignment of the rods by applied electric fields, fluid flow, or substrates in order
to facilitate the formation of the desired self-assembled structures \cite{baranov}. Our simulations show that
this additional "steering" is required since the spontaneous nucleation of the rods is strongly affected by glassy
dynamics and spinodal instabilities.

\begin{acknowledgments}
Financial support of a NWO-VICI grant is acknowledged. This work is part of the research programme of FOM, which is financially supported by NWO.
\end{acknowledgments}

\end{document}